\documentclass[a4paper,11pt]{article}

\newcommand{\sla}[1]{#1\kern-0.55em/}
\usepackage{jheppub}
\usepackage{graphicx,epsfig}
\usepackage{dcolumn}
\usepackage{bm}
\usepackage{amsmath}
\usepackage{amssymb}
\usepackage{float}
\renewcommand\thefootnote{\fnsymbol{footnote}}

\begin{document}

\hfill {\tt  KCL-PH-TH/2019-90, CERN-TH-2019-211}  

\def\thefootnote{\fnsymbol{footnote}}
 
\begin{center}

\vspace{2.cm}

{\Large\bf Exploring Supersymmetric CP Violation after LHC Run~2 \\
\vspace{0.2cm}
with Electric Dipole Moments and B Observables}

\setlength{\textwidth}{11cm}
                    
\vspace{1cm}
{\large\bf  
A.~Arbey$^{\,a,b,}$\footnote{Also Institut Universitaire de France, 103 boulevard Saint-Michel, 75005 Paris, France}$^{,}$\footnote{Email: alexandre.arbey@ens-lyon.fr}, 
J.~Ellis$^{b,c,d,}$\footnote{Email: John.Ellis@cern.ch},
F.~Mahmoudi$^{a,b,*,}$\footnote{Email: nazila@cern.ch}
}
 
\vspace{0.5cm}
{\em $^a$Universit\'e de Lyon, Universit\'e Claude Bernard Lyon 1, CNRS/IN2P3, \\
Institut de Physique des 2 Infinis de Lyon, UMR 5822, F-69622, Villeurbanne, France}\\[0.2cm]
{\em $^b$Theoretical Physics Department, CERN, CH-1211 Geneva 23, Switzerland} \\[0.2cm]
{\em $^c$Theoretical Particle Physics and Cosmology Group, Department of Physics,\\
King's College London, London WC2R 2LS, United Kingdom}\\[0.2cm] 
{\em $^d$National Institute of Chemical Physics \& Biophysics, R{}\"avala 10, 10143 Tallinn, Estonia}\\[0.2cm]

\end{center}

\renewcommand{\thefootnote}{\arabic{footnote}}
\setcounter{footnote}{0}

\vspace{0.5cm}
\thispagestyle{empty}
\centerline{\bf ABSTRACT}
\vspace{0.5cm}
We consider the prospects for measuring distinctive signatures of the CP-violating
phases in the minimal supersymmetric extension of the Standard Model (MSSM) in light
of the limits on sparticle masses from searches at the LHC. We use the {\tt CPsuperH} code to evaluate
model predictions and scan the parameter space using a geometric approach that maximizes 
CP-violating observables subject to the current upper limits
on electric dipole moments (EDMs). We focus on the possible CP-violating asymmetry $A_{\rm CP}$ in 
$b \to s \gamma$ decay and on a possible CP-violating contribution to the 
$B_s - \overline{B}_s$ mass difference $\Delta M^{NP}_{B_s}$, as well as future measurements 
of the EDMs of the proton, neutron and electron. We find that the current LHC and EDM limits
are consistent with values of  $A_{\rm CP}$, $\Delta M^{NP}_{B_s}$ and the proton EDM that are
measurable with the Belle-II detector, LHCb and a proposed measurement of the proton EDM
using a storage ring, respectively. Measurement of a non-zero proton EDM would constrain $A_{\rm CP}$
significantly, but it and a CP-violating contribution to $\Delta M^{NP}_{B_s}$ could still be measurable,
along with neutron and electron EDMs. A more accurate measurement of $A_{\rm CP}$
with the current central value would favour stop and chargino masses within reach of future LHC runs
as well as a potentially measurable value of $\Delta M^{NP}_{B_s}$.


\newpage

\section{Introduction}
\label{sec:intro}

The lack of any signal for supersymmetry during Run 2 of the LHC has triggered premature
gloom in some quarters. There is still considerable scope for supersymmetry to be hiding shyly in
some more subtle channels than those studied so far, or at the higher masses that will be
explored with over an order of magnitude more data during future LHC runs. Moreover, as
we discuss in this paper, there is also scope for distinctive signatures of the CP-violating phases allowed in the minimal supersymmetric extension of the Standard Model (MSSM), despite
the continuing success of the Kobayashi-Maskawa model \cite{PDG}.

There are six potentially observable CP-violating phases in a phenomenological version of the 
MSSM with soft supersymmetry-breaking parameters specified at the electroweak scale in which
flavour universality is assumed for the supersymmetric partners of the two lighter generations
(the pMSSM). This maximally CP-violating, minimally flavour-violating (MCPMFV) model~\cite{Ellis:2007kb}
contains the three phases of the SU(3), SU(2) and U(1) gaugino masses, 
$\phi_{M_3,M_2,M_1}$, and the phases of the third-generation trilinear soft supersymmetry-breaking 
parameters, $\phi_{A_t,A_b,A_\tau}$~\footnote{We do not consider the phases of the trilinear
parameters for the first and second generations, which have much less phenomenological
impact. We also assume that the CP-violating QCD phase is negligible.}, which we allow to vary
independently. 

The experimental upper limits on electric dipole moments (EDMs) compiled in
Table~\ref{tab:EDMs} constrain quite severely four combinations of these six pMSSM CP-violating
phases. However, they leave open the possibility that two linearly-independent orthogonal
combinations of these six phases may take values that are ${\cal O}(1)$, in which case they may
have observable consequences in other experiments, such as $B$-physics experiments (see below) as well as in the next generation of
EDM experiments.

\begin{table}[!ht]
\begin{center}
\begin{tabular}{|c|c|c|}
 \hline
 Element/Particle & Upper limit (e.cm) & Reference\\
 \hline\hline
 Thallium (2002) & $1.1\times10^{-24}$  & \cite{Regan:2002ta}\\
 \hline
 Muon (2008) & $1.8\times10^{-19}$ & \cite{Bennett:2008dy}\\
 \hline
 Mercury (2016) & $7.4\times10^{-30}$ & \cite{Graner:2016ses}\\
 \hline
 Thorium Monoxide (2018) & $1.3\times10^{-29}$ & \cite{Andreev:2018ayy} \\
 \hline
 Neutron (2020) & $2.2\times10^{-26}$ & \cite{Abel:2020gbr}\\
 \hline\hline
 Proton (future) & $5\times10^{-29}$ & \cite{Anastassopoulos:2015ura} \\
 \hline
\end{tabular}
\caption{\it The 95\% C.L. upper limits on EDMs used as constraints in this study.}
\label{tab:EDMs}
\end{center}
\end{table}

When exploring these CP-violating opportunities, it is impractical to make an analytical
study of the possible phase values, and a random scan of the multidimensional parameter
space is inefficient. A geometric approach to this type of problem was proposed in~\cite{Geometry}, which was used in~\cite{Ellis:2010xm} to analyze the impacts in certain benchmark MSSM scenarios
of three EDM constraints. This approach was adapted in~\cite{Arbey:2014msa}
to analyze the constraints imposed on the MCPMFV scenario by a set of four EDM measurements.
This paper also extended the approach beyond the small-phase approximation,
and we follow a similar strategy in this paper. As in~\cite{Arbey:2014msa}, the tool we use to
calculate the effects of the CP-violating pMSSM phases is the {\tt CPsuperH} code~\cite{Lee:2003nta,Lee:2007gn,Lee:2012wa}.

There have been two important experimental developments since the previous analysis~\cite{Arbey:2014msa}.
One is the publication of significant improvements in the upper limits on a pair of EDMs: {as seen in Table~\ref{tab:EDMs},} the
upper limit on the EDM of Thorium Monoxide has improved by an order of magnitude, and that
on the EDM of Mercury has improved by a factor approaching 5. The other important
experimental development has been the strengthening of lower limits on sparticle masses
following unsuccessful searches for sparticles during Run~2 of the LHC. The interpretation
of these data is quite complex, being dependent on details of the sparticle mass spectrum such
as the degrees of degeneracy between different sparticle species, etc. It was found {in~\cite{Athron:2017yua,Bagnaschi:2017tru}}
that strongly-interacting sparticles could be significantly lighter than the values often quoted on
the basis of analyses of missing-energy signatures in simplified models. The more relaxed
sparticle mass limits assumed here, which are motivated by the analyses in~\cite{Athron:2017yua,Bagnaschi:2017tru},
are listed in Table~\ref{tab:LHCmass}. In addition to these constraints, we also apply the LEP constraints on electroweakino and slepton masses compiled in~\cite{Tanabashi:2018oca}.

\begin{table}[!ht]
\begin{center}
\begin{tabular}{|c|c|}
 \hline
 Particle & Mass limit (GeV)\\
 \hline\hline
 Lightest Higgs boson & $\in [122,128]$ \\
\hline
 Gluino & $> 2000$\\
 \hline
 First- and second-generation squarks & $> 1000$\\
 \hline
 Third-generation squarks & $> 500$\\
 \hline
\end{tabular}
\caption{\it The LHC mass limits applied here, which are motivated by the {analyses in~\cite{Athron:2017yua,Bagnaschi:2017tru}}.\label{tab:LHCmass}}
\end{center}
\end{table}

As already mentioned, we consider three types of prospective CP-violating observables.
One is the CP-violating asymmetry $A_{\rm CP}$ in $b \to s \gamma$ decay, which is
currently measured {to be $0.015 \pm 0.011$ \cite{Tanabashi:2018oca}}, and may be measured by Belle-II
with a precision of {$\pm \, 0.002$~\cite{Kou:2018nap}}. The second observable is the possible new-physics
contribution to $B_s - \overline{B}_s$ mass difference, $\Delta M^{NP}_{B_s}$, which is
currently constrained to be $< 3.1$/ps at 95\% C.L. \cite{Tanabashi:2018oca,Dowdall:2019bea}. 
The most important limitation on this constraint
is due to the accuracy of theoretical calculations (principally using lattice QCD) of the 
hadronic matrix element that controls the magnitude of the Standard Model contribution.\footnote{For a short review of the current status of the non-perturbative determination of the $B$-mixing parameters we refer the reader to~\cite{DiLuzio:2019jyq} and the references therein.} 
We assume that this uncertainty can be reduced sufficiently for a determination by LHCb
with a precision of $\pm 0.04$/ps to be possible. Finally, we consider prospective
measurements of EDMs.

As well as improvements in the sensitivities to the `classic' electron and neutron EDMs,
a new possibility is a measurement of the EDM of the proton using a storage ring,
which has a prospective sensitivity of $\pm \, 0.025 \times 10^{-27}$~e.cm{~\cite{Anastassopoulos:2015ura},
see also~\cite{PBC}.} In order to
assess the power of this measurement, we consider two scenarios: one in which the 
central value is zero, and one in which the central value of the proton EDM is
$1 \times 10^{-27}$~e.cm.

The outline of our paper is as follows. In Section~\ref{sec:method} we review the
geometric method we use to analyze efficiently the possibilities left open by the
new generation of constraints on the MCPMFV model. Then, in Section~\ref{sec:pMSSM}
we apply this method to the MCPMFV extension of the pMSSM, considering the current constraints,
the potential implications of a future proton EDM measurement, and also those of a more precise
measurement of $A_{\rm CP}$ with the current central value. Finally, in
Section~\ref{sec:conclusions} we summarize our conclusions.

\section{Geometric Sampling Method}
\label{sec:method}

\subsection{Small-Phase Expansion}

We consider initially the constraints imposed on the six MCPMFV phases ${\bf \Phi} \equiv \Phi_\alpha = \Phi_{1,2,3,t,b,\tau}$
by the four most important EDM measurements $E^i$ in the small-phase approximation. In this case, one may write
\begin{equation}
E^i \; \simeq \; {\mathbf \Phi}.{\mathbf E}^i \, ,
\label{smallangle}
\end{equation}
where ${\mathbf E}^i \equiv \partial E^i/\partial {\mathbf \Phi}$,
i.e., $E^i_\alpha \equiv \partial E^i/\partial \Phi_\alpha$. The index $i$ therefore refers to observables and the subscript $\alpha$ to the space of phases.
The EDM vectors $E^i_\alpha$ span a  codimension-four subspace of the six-dimensional space of CP-violating
phases $\Phi_\alpha$ that is defined by the following quadruple exterior product:
\begin{equation}
 A_{\alpha\beta\gamma\delta} \; = \; E^a_{[ \alpha} \, E^b_\beta \, E^c_\gamma \, E^d_{\delta ]} \, ,
\label{subspace}
\end{equation}
where the symbols $[ ... ]$ denote antisymmetrized indices. This subspace is a
two-dimensional plane. In general, the dependence on the phases $\Phi_\alpha$
of a generic CP-violating observable $O$ in the
small-phase approximation is given by ${\mathbf O} \equiv \partial O/\partial {\mathbf \Phi}$,
i.e., $O_\alpha \equiv \partial O/\partial \Phi_\alpha$. The six-vector
\begin{equation}
 B_\mu \; \equiv \; \epsilon_{\mu\nu\lambda\rho\sigma\tau}\, O_\nu \, E^a_\lambda \, E^b_\rho\, E^c_\sigma \, E^d_\tau
\end{equation}
characterizes a direction in the space of CP-violating phases where neither the EDMs nor
the observable $O$ receive contributions in the small-phase approximation. The direction in the space of
CP-violating phases that optimizes $O$ while not contributing to the EDMs is
orthogonal to $B_\mu$ as well as to the EDM vectors $E^{a,b,c,d}_\alpha$, and is characterized
by the six-vector
\begin{eqnarray}
 \Phi_\alpha & = & \epsilon_{\alpha\beta\gamma\delta\mu\eta} \, E^a_\beta \, E^b_\gamma \, E^c_\delta \, E^d_\mu \, B_\eta \nonumber \\
 & = & \epsilon_{\alpha\beta\gamma\delta\mu\eta} \, \epsilon_{\eta\nu\lambda\rho\sigma\tau} \, E^a_\beta \, E^b_\gamma \, E^c_\delta \, E^d_\mu \, O_\nu \, E^a_\lambda \, E^b_\rho \, E^c_\sigma \, E^d_\tau \, ,
\label{optimal}
\end{eqnarray}
where the normalization factor is arbitrary.

\subsection{Extension to Larger Phases}

The linear geometric approach used in~\cite{Ellis:2010xm} entailed first
fixing the phases to $0^\circ$ or $\pm180^\circ$ for each choice of MSSM parameters to be scanned, next
identifying the optimal direction for an observable $O$ of interest using the above geometric approach, and then
choosing randomly sets of phases along this direction. As was discussed in~\cite{Arbey:2014msa}, this procedure
requires modification when the phases are no longer small. We follow here the iterative approach 
suggested in~\cite{Arbey:2014msa} that extends and improves the efficiency of the linear geometric approach.
After using this approach to compute the optimal direction for initial phases $0^\circ$ or $\pm180^\circ$,
we next move by $20^\circ$ in this direction, and then
recompute the optimal direction at this new point and iterate up to $100^\circ$.

\subsection{Scanning Strategy}

As mentioned in the Introduction, the MCPMFV contains six phases, $\phi_{M_3,M_2,M_1}$ and $\phi_{A_t,A_b,A_\tau}$ in addition to the 19 standard pMSSM parameters: the gaugino and Higgsino masses $M_{1,2,3}$ and $\mu$, the squark masses $M_{Q_L,Q_{3L},D_R,b_R,U_R,t_R}$ and trilinear couplings $A_{b,t}$, the sleptons masses $M_{L_L,\tau_L,L_R,\tau_R}$ and trilinear coupling $A_\tau$, charged Higgs mass $M_{H^\pm}$ and $\tan\beta$. We consider gluino and squark masses between 500 GeV and 5 TeV, moduli of the squark trilinear couplings below 12 TeV, the other masses between 50 GeV and 3 TeV, $|A_\tau| < 7$ TeV and $\tan\beta \in [2,60]$.

We performed random flat scans over the CP-conserving pMSSM parameters and use the above geometric approach for the CP-violating phases. We have generated several million points in our study of the MCPMFV extension of the pMSSM,
and retained only those points yielding a neutral Higgs boson
with mass in the range $122-128$ GeV, which corresponds to the measured value $m_h \simeq 125$~GeV
within a conservative theoretical uncertainty.~\footnote{We have checked that varying the Higgs mass range between $\pm 1$ and $\pm 5$~GeV does not affect our results, apart from changing the statistics.} In addition, we require the LSP to be the lightest neutralino.

We use the {\tt CPsuperH}~\cite{Lee:2003nta,Lee:2007gn,Lee:2012wa} code~\footnote{We have verified for a subset of relevant parameter
sets that the Higgs mass calculated with the {\tt FeynHiggs}~\cite{FH} and {\tt SPheno}~\cite{SPh} codes are mostly
within $\pm 1$~GeV of the {\tt CPsuperH} values,  well within the mass range we use.}
to calculate the MCPMFV pMSSM mass spectra and couplings, as well as the EDM constraints, apart from that for
thorium monoxide, for which we use~\cite{Cheung:2014oaa}.
We use the {\tt SuperIso}~\cite{Mahmoudi:2007vz,Mahmoudi:2008tp,Mahmoudi:2009zz} and {\tt CPsuperH} codes to calculate flavour constraints. Also, we have used {\tt SuperIso Relic}~\cite{Arbey:2009gu,Arbey:2011zz,Arbey:2018msw} to calculate dark matter observables. However, we found that dark matter constraints do not affect the CP phases, and that their main effects are similar to those in the CP-conserving pMSSM \cite{Arbey:2017eos}. In addition, we use {\tt HiggsBounds}~\cite{Bechtle:2008jh,Bechtle:2011sb,Bechtle:2013wla} to compare model predictions with the LHC heavy Higgs constraints. We also found that heavy Higgs searches do not affect the CP properties of the pMSSM. Finally, we require that the lightest Higgs couplings are consistent with the latest ATLAS experimental measurement combination \cite{Aad:2019mbh} at the 95\% C.L.

\subsection{Sparticle Masses in the Scan}

\begin{figure}[t!]
\begin{center}
\begin{tabular}{cc}
\includegraphics[width=0.33\paperwidth]{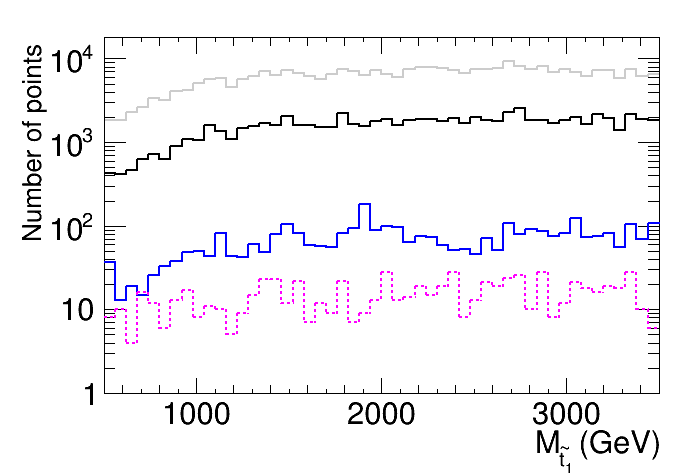} & 
\includegraphics[width=0.33\paperwidth]{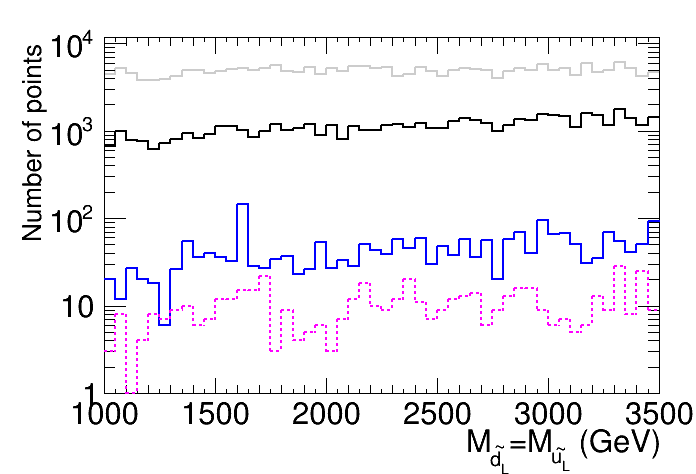} \\
\includegraphics[width=0.33\paperwidth]{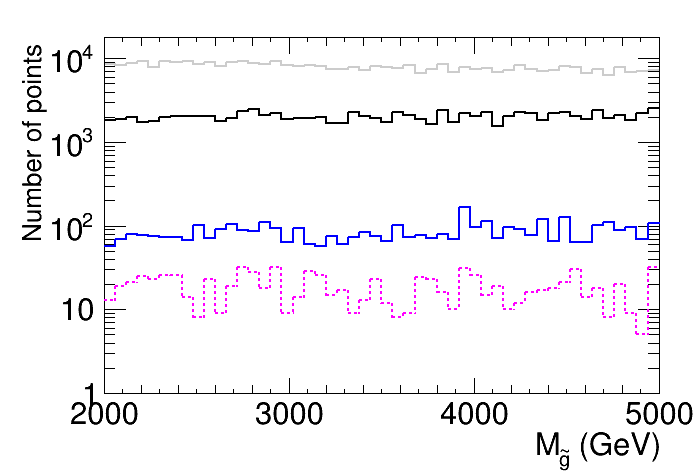} & 
\includegraphics[width=0.33\paperwidth]{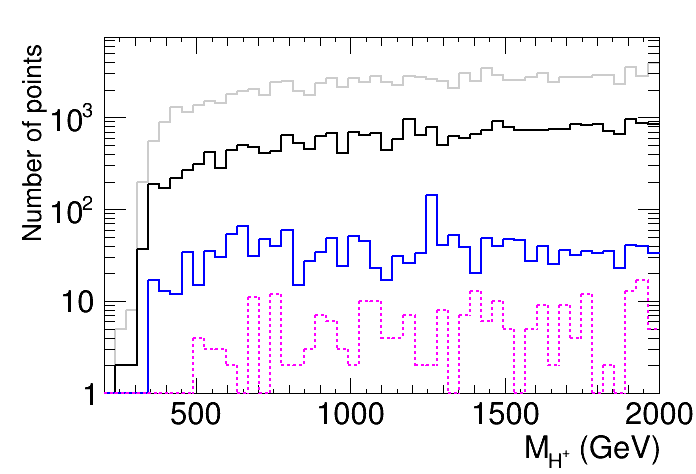}\\
 \end{tabular}
\caption{\it Distributions of sparticle masses in our scan of the MCPMFV pMSSM.
Upper left: The lighter stop mass, upper right: left-handed first- and second-generation
squark mass, lower left: gluino mass, lower right: charged Higgs mass.
Grey histograms: all points passing the LHC mass and flavour constraints. Black: also including current EDM limits; blue: including possible future proton EDM constraint, assuming a central value of zero, magenta: including possible future proton EDM measurement of $(1\pm0.025)\times10^{-27}$ e.cm.} 
\label{masses1}
\end{center}
\end{figure}

\begin{figure}[ht!]
\begin{center}
\begin{tabular}{cc}
\includegraphics[width=0.33\paperwidth]{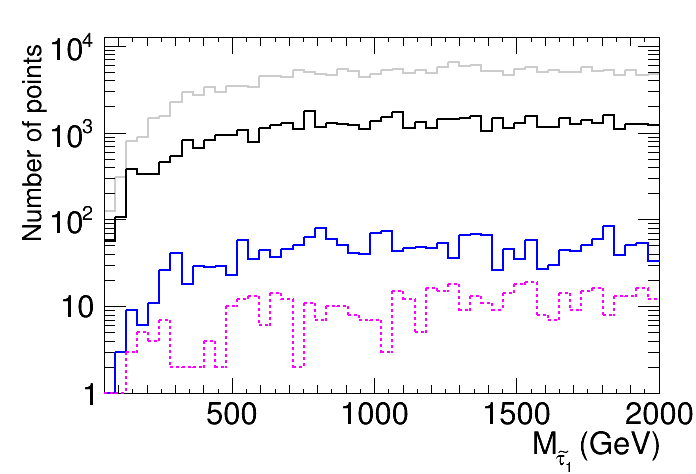} & 
\includegraphics[width=0.33\paperwidth]{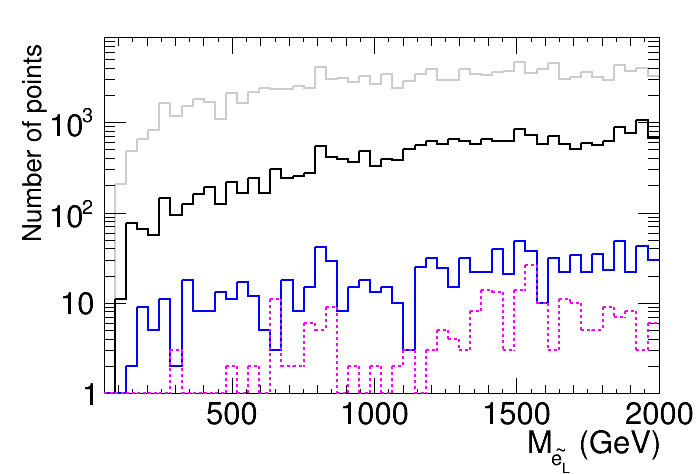} \\
\includegraphics[width=0.33\paperwidth]{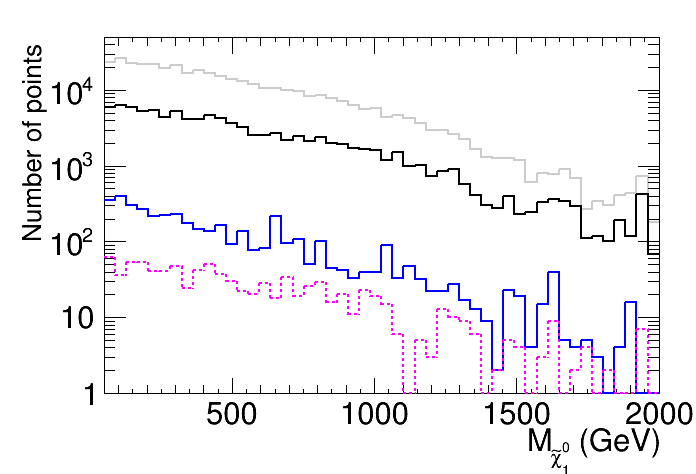} & 
\includegraphics[width=0.33\paperwidth]{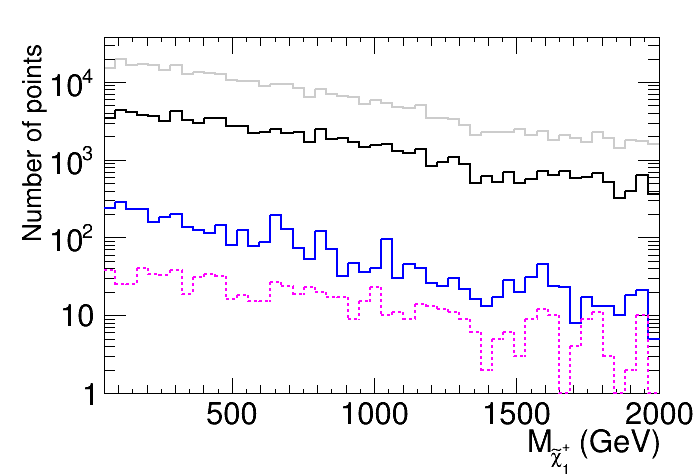}\\
 \end{tabular}
\caption{\it Distributions of sparticle masses in our scan of the MCPMFV pMSSM (continued).
Upper left: The lighter stau mass, upper right: left-handed selectron mass, lower left: lightest neutralino mass, lower right: lighter chargino mass.
Grey histograms: all points passing the LEP and LHC mass and flavour constraints. Black: also including current EDM limits; blue: including possible future proton EDM constraint, assuming a central value of zero, magenta: including possible future proton EDM measurement of $(1\pm0.025)\times10^{-27}$ e.cm.} 
\label{masses2}
\end{center}
\end{figure}

We assume the sparticle mass limits shown in Table~\ref{tab:LHCmass}, namely that the gluino
weighs $\ge 2000$~GeV, first- and second-generation squarks weigh $\ge 1000$~GeV and
third-generation squarks weigh $\ge 500$~GeV. These mass limits are motivated by the analyses
of the CP-conserving pMSSM in~\cite{Athron:2017yua,Bagnaschi:2017tru}, which included constraints from LHC 13-TeV data. Figures~\ref{masses1} and \ref{masses2} illustrate the distributions of sparticle masses retained in our scan. Here and in subsequent figures, the grey histograms include all the points passing the LHC mass and flavour constraints, the black histograms also include the current EDM limits, the blue histograms include a possible future proton EDM constraint assuming a null measurement with a central value of zero, and the magenta histograms include a possible future discovery of a proton EDM of $(1\pm0.025)\times10^{-27}$ e.cm~\cite{Anastassopoulos:2015ura}.

The upper left panel of Fig.~\ref{masses1} shows the distribution of the lighter stop mass, $M_{\tilde t_1}$, to which that of the lighter bottom squark mass is very similar, and the upper right panel shows that of the left-handed up- and down-flavoured squarks, which are again very similar, $M_{\tilde u_L} \simeq M_{\tilde d_L}$, as are those of the right-handed up- and down-flavoured squarks. The lower left panel shows the distribution of the gluino mass, $M_{\tilde g}$, and the lower right panel shows that of the charged Higgs bosons, $M_{H^\pm}$, to which the masses of the two heavier neutral Higgs bosons are quite similar. 
We see that these mass distributions are all quite flat, with stop and  $H^\pm$ masses away from the lower limits even being slightly favoured. These observations give us some confidence that our results are not biased by the approximate nature of our assumed lower bounds, and that they may be robust with respect to near-future improvements in the LHC mass reaches.

Figure~\ref{masses2} shows the distributions of the masses of some electroweakly-interacting sparticles, the lighter stau (upper left panel), the left-handed selectron (upper right panel, results for the right-handed selectron are very similar), the lightest neutralino, which we assume to be the lightest supersymmetric particle and provide the astrophysical cold dark matter (lower left panel), and the lighter chargino (lower right panel). Our scan did not include explicitly lower LHC limits on these masses, which are known to be dependent on the details of the sparticle spectrum, but we do impose the LEP constraints on the electroweakino and slepton masses~\cite{Tanabashi:2018oca}. We note that our slepton spectra are peaked at higher masses, providing more reassurance of the robustness of our results. 

We checked the impact of the Higgs mass and flavour constraints, and found that both constraints affect the shape of the grey LHC histograms for small masses. The shapes of the neutralino and chargino mass distributions are due, however, to the random scanning approach and the requirement that the LSP be a neutralino, while the distribution of the phases are due to the geometric approach, which favours small phases.
 
\subsection{Prospective Measurement of the Proton EDM}

One of the foci of our analysis is the prospective impact of a measurement of the EDM of the proton. There is a proposal to construct a storage ring at CERN with a prospective uncertainty of $\pm \, 0.025 \times10^{-27}$ e.cm~\cite{Anastassopoulos:2015ura}, almost three orders of magnitude smaller than the uncertainty in the current null measurement of the neutron EDM. Significant improvements in neutron EDM measurements are envisaged in several experiments, but we use the prospective proton EDM measurement as a benchmark for the future experimental sensitivity to hadron EDMs.

\begin{figure}[ht!]
\begin{center}
\begin{tabular}{cc}
\includegraphics[width=0.33\paperwidth]{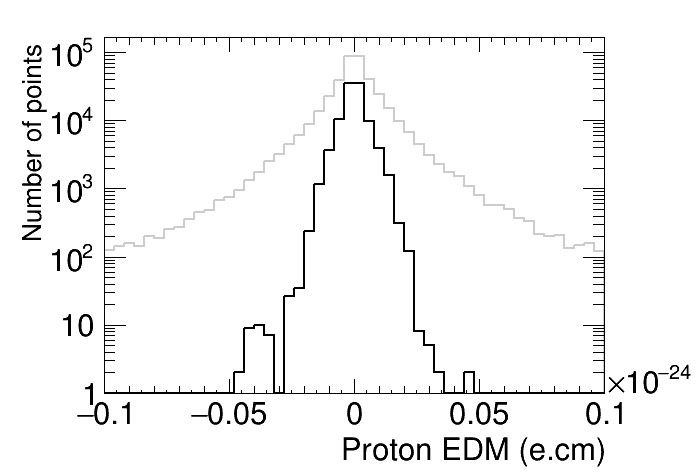} & 
\includegraphics[width=0.33\paperwidth]{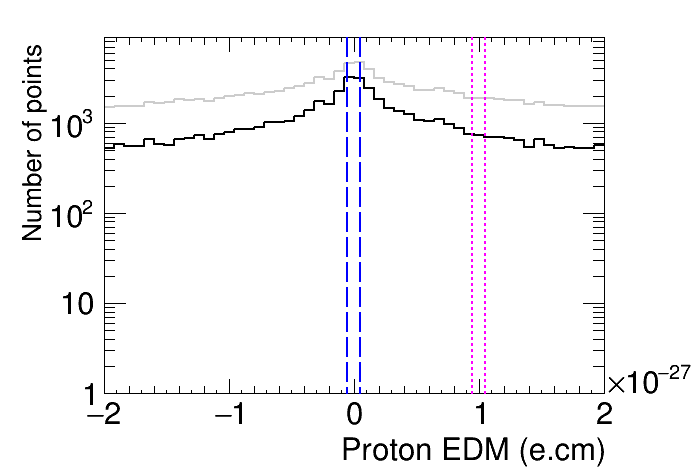}
 \end{tabular}
\caption{\it Distributions of the proton EDM, showing the effects of taking into account the LHC constraints on sparticle masses and imposing the current EDM constraints. Gray histograms: all points passing LHC mass and flavour constraints; black: including current EDM limits. The right plot is a zoom, where the dashed vertical lines show the prospective proton EDM measurements that we consider: either $(0\pm0.025)\times10^{-27}$ e.cm (blue dashed lines) or $(1\pm0.025)\times10^{-27}$ e.cm (purple dashed lines).} 
\label{pEDM}
\end{center}
\end{figure}

As already mentioned, we consider two hypotheses for the result of such an experiment: a null measurement with a central value of zero, and a possible future discovery of a proton EDM of $(1\pm0.025)\times10^{-27}$ e.cm. The left panel of Fig.~\ref{pEDM} displays histograms from our scan of the possible value of the proton EDM imposing the LHC mass constraints discussed above and all the current flavour constraints (gray) and including also the current EDM constraints (black). Unsurprisingly, these restrict severely the possible range of the proton EDM.  
The right panel of Fig.~\ref{pEDM} displays a zoomed view of values of the proton EDM $\in [-2, 2] \times 10^{-27}$ e.cm. The two sets of vertical dashed lines correspond to our two hypotheses for the future measurement: the null hypothesis (blue) and the discovery hypothesis (magenta). The impacts of these two possible hypotheses are discussed in detail in the next Section, but we note here that the shapes of the mass histograms in Figs.~\ref{masses1} and \ref{masses2} are not very sensitive to the specific proton EDM hypothesis assumed.

For the purpose of assessing the LHC constraint projections that can be anticipated on the time scale of the proton EDM projection, we increased the lower limits on the sparticle masses and decreased the uncertainties on the lightest Higgs coupling constraints. However, beyond a decrease in statistics and cuts in the mass distributions, the results are not affected significantly, {\it i.e.}, the distribution shapes remain unchanged. Therefore, in order to maximize the statistics, we use the current experimental constraints in the following.

\section{Analysis of CP-Violating Effects}
\label{sec:pMSSM}

\subsection{CP-Violating Phases}

\begin{figure}[t!]
\begin{center}
\begin{tabular}{cc}
\includegraphics[width=0.33\paperwidth]{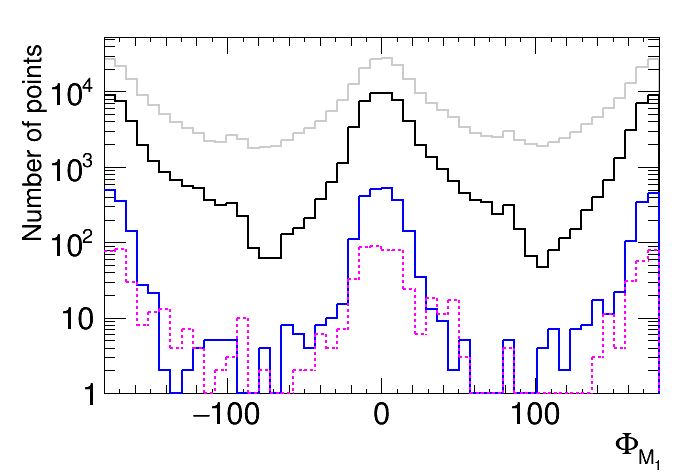} & 
\includegraphics[width=0.33\paperwidth]{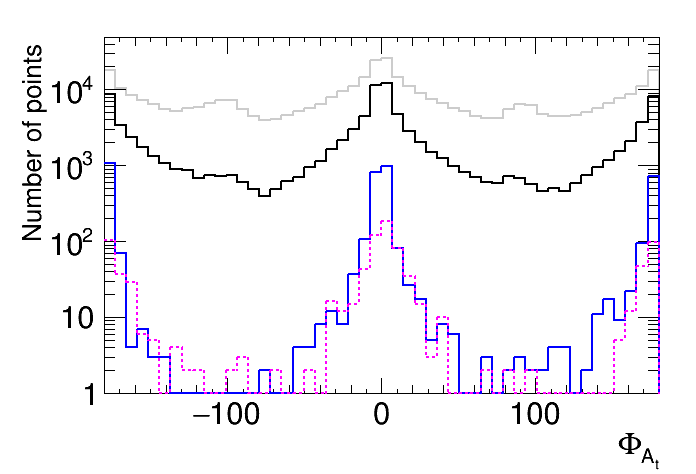}\\
\includegraphics[width=0.33\paperwidth]{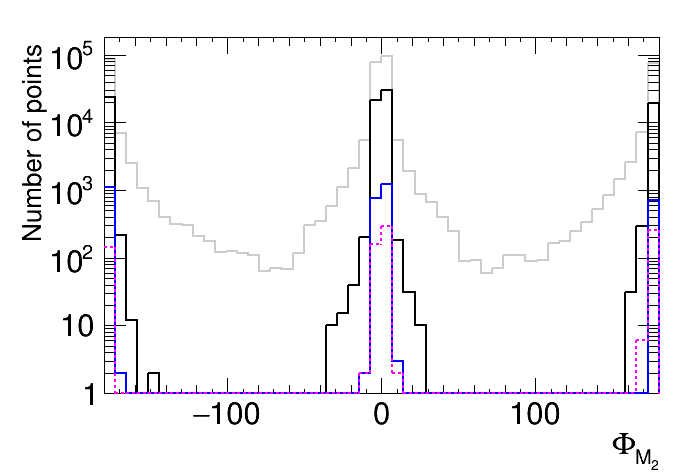}&
\includegraphics[width=0.33\paperwidth]{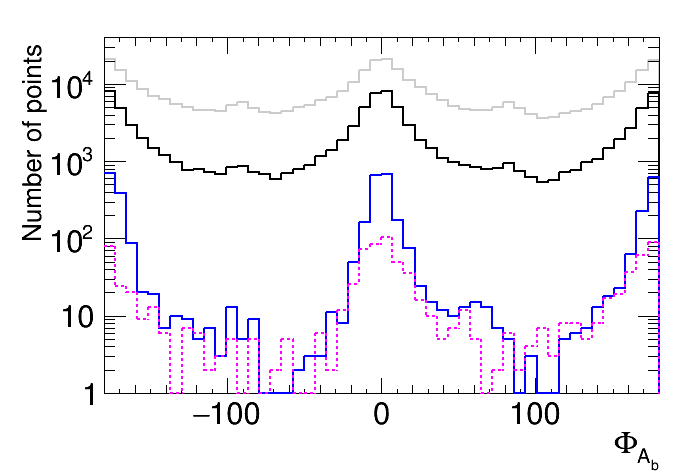} \\ 
\includegraphics[width=0.33\paperwidth]{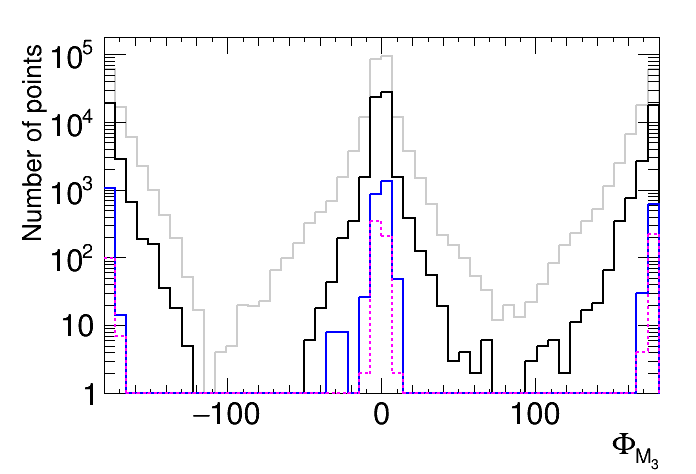} & 
\includegraphics[width=0.33\paperwidth]{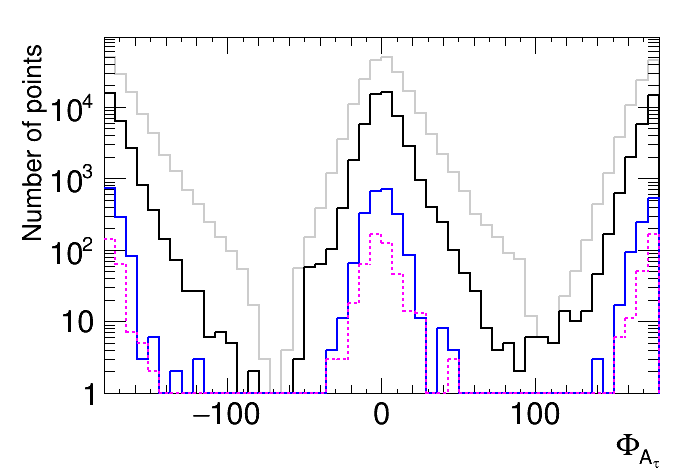}
 \end{tabular}
\caption{\it Distributions of CP-violating phases, showing the effects of taking into account the LHC constraints on sparticle masses and imposing the EDM constraints. Left column, from top to bottom: the gaugino phases $\phi_{M_1},\phi_{M_2},\phi_{M_3}$; right column, from top to bottom: the trilinear phases $\phi_{A_t},\phi_{A_b},\phi_{A_\tau}$.
Grey histograms: all points passing the LHC mass and flavour constraints; black: also including current EDM limits; blue: including possible future proton EDM constraint, assuming a central value of zero, magenta: including possible future proton EDM measurement of $(1\pm0.025)\times10^{-27}$ e.cm.} 
\label{phases}
\end{center}
\end{figure}

As discussed above, the four EDM constraints each impose some combination of
CP-violating phases in the MCPMFV pMSSM to be small, but are not sufficient to
force all of its six phases to be small simultaneously. This is seen explicitly in
Fig.~\ref{phases}, where we display histograms of the values of the phases found
in our optimized parameter scan. The values of the gaugino phases $\phi_{M_{1,2,3}}$
are shown in the left column of panels, and the values of the phases of the 
third-generation trilinear phases are shown in the right column.  As could be expected,
when the current flavour - but not EDM - constraints are imposed (gray histograms),
large values of the CP-violating phases are all relatively easy to obtain. When the current
EDM constraints are also imposed (black histograms), we still find large values of
$\phi_{M_1}, \phi_{A_\tau}$ and $\phi_{A_b}$ relatively easily, whereas large values of
$\phi_{A_t}$ are rarer though values $\sim \pi/2$ are possible, and values of 
$\phi_{M_{2,3}} \sim \pm \pi/2$ are absent.

As could be expected, when the null hypothesis for a future proton EDM measurement 
is assumed (blue histograms), the histograms for $\phi_{M_1}, \phi_{A_t}, \phi_{A_\tau}$ and $\phi_{A_b}$
are all depleted at values $\sim \pm \pi/2$, though CP-violating phases as large as 
several tens of degrees are still possible. However, {this is not possible} for
$\phi_{M_2}$ and $\phi_{M_3}$. Interestingly, the histograms for the non-zero
hypothesis for the proton EDM (magenta) are quite similar, making the point that
any proton EDM measurement with the estimated precision $\pm \, 0.25 \times 10^{-27}$ e.cm
would be comparably restrictive.

In general, measurements of the electron EDM impose strong constraints on the electroweak sector, e.g., a combination of the $M_2$ and $A_\tau$ phases, and a precise measurement of the proton EDM will allow us to probe in addition the strong sector, e.g., a combination of the $M_3$, $A_t$ and $A_b$ phases. On the other hand, the $M_1$ phase will only very weakly be affected by the EDM measurements because the bino mass only appears indirectly in the neutralino mixing matrix.

\begin{figure}[ht!]
\begin{center}
\begin{tabular}{cc}
\includegraphics[width=0.33\paperwidth]{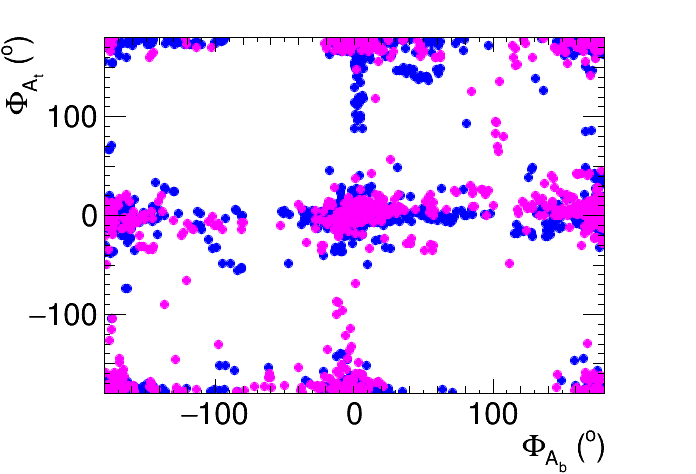} & 
\includegraphics[width=0.33\paperwidth]{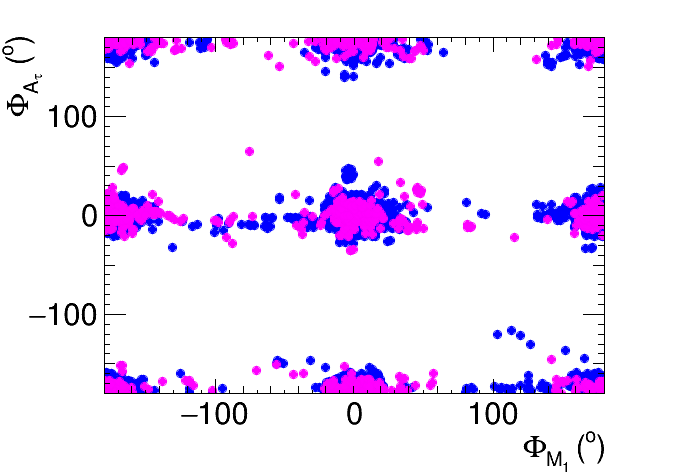}
 \end{tabular}
\caption{\it Distributions of scan points in the $(\phi_{A_b},\phi_{A_t})$ (left) and $(\phi_{M_1},\phi_{A_\tau})$ (right) parameter planes, showing the effects of taking into account the LHC constraints on sparticle masses and imposing the EDM constraints. Blue:  including possible future proton EDM constraint, assuming a central value of zero; magenta: including possible future proton EDM measurement of $(1\pm0.025)\times10^{-27}$ e.cm.} 
\label{phaseplanes}
\end{center}
\end{figure}

Some examples of correlations between phases are shown in Fig.~\ref{phaseplanes} in two planes, one with strong-sector phases and the other with electroweak-sector phases. 
The left panel shows the $(\phi_{A_b},\phi_{A_t})$ plane and the right panel shows the
$(\phi_{M_1},\phi_{A_\tau})$ plane, and in each panel the case of the null proton EDM
hypothesis is illustrated by blue dots and the discovery hypothesis by magenta dots.
We see again in the left panel that large values of $\phi_{A_b}$ and $\phi_{A_t}$
are possible, and that they may have large values simultaneously if their signs are
correlated, as indicated by the points in the upper right and lower left quadrants. On the
other hand, we see again in the right panel that, while large values of $\phi_{M_1}$ are
possible, large values of $\phi_{A_\tau}$ are disfavoured. 

While cases with one large phase are relatively easy to find and correspond in general to a large value of the corresponding sparticle mass, scenarios in which two (or more) phases are large are much less common, because they are the results of accidental cancellation between the different terms contributing to the proton EDM.

\subsection{EDMs}

\begin{figure}[t!]
\begin{center}
\includegraphics[width=0.4\paperwidth]{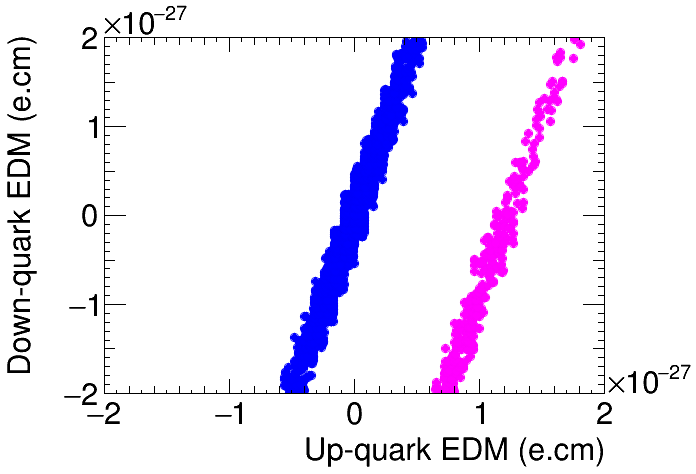}
\caption{\it Distributions of scan points in the up-quark EDM vs. down quark EDM plane, showing the effects of taking into account the LHC constraints on sparticle masses and imposing the EDM constraints. Blue:  including possible future proton EDM constraint, assuming a central value of zero; magenta: including possible future proton EDM measurement of $(1\pm0.025)\times10^{-27}$ e.cm.} 
\label{qEDMs}
\end{center}
\end{figure}

We now explore the implications of a proton EDM measurement for other EDMs.
As seen in Fig.~\ref{qEDMs}, the prospective precision of the proton EDM measurement
would enforce a strong correlation between the up- and down-quark EDMs. However, this 
would be different in the case of a null measurement, when it would be centred around zero, 
from the case of the discovery hypothesis, in which case an up-quark EDM of ${\cal O}(10^{-27})$ e.cm would be favoured. In both cases, values of the down-quark larger in magnitude than 
${\cal O}(2 \times 10^{-27})$ e.cm would be quite possible.

\begin{figure}[t!]
\begin{center}
\begin{tabular}{cc}
\includegraphics[width=0.33\paperwidth]{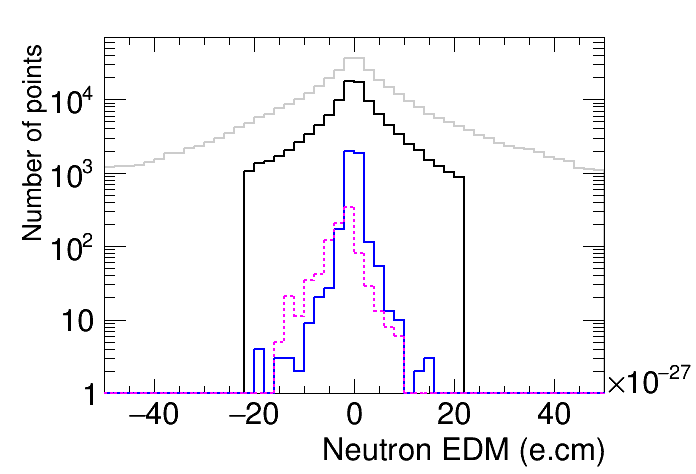} & 
\includegraphics[width=0.33\paperwidth]{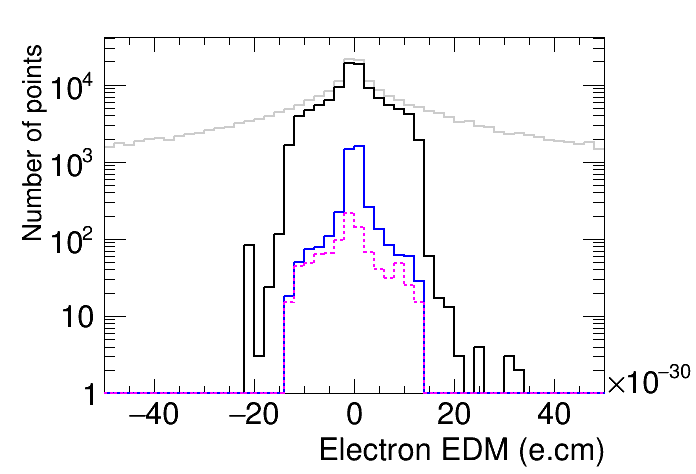}
 \end{tabular}
\caption{\it Distributions of the neutron EDM (left) and the electron EDM (right), showing the effects of taking into account the LHC constraints on sparticle masses and imposing the EDM constraints. Grey histogram: all points passing the LHC mass and flavour constraints; black: also including current EDM limits; blue: including possible future proton EDM constraint, assuming a central value of zero; magenta: including possible future proton EDM measurement of $(1\pm0.025)\times10^{-27}$ e.cm.}
\label{peEDMs}
\end{center}
\end{figure}

This observation is reflected in the left panel of Fig.~\ref{peEDMs}, where we see that values
of the neutron EDM that are larger in magnitude than $10^{-26}$ e.cm were found in our scan,
in both the proton EDM hypotheses. Moreover, even the null hypothesis for the proton EDM would not
exclude the possibility that the neutron EDM is over two orders of magnitude larger. The
right panel of Fig.~\ref{peEDMs} tells a similar story for the electron EDM. The present EDM
limits (black histogram) already constrain the electron EDM to be $< {\cal O}(2 \times 10^{-29})$ e.cm in magnitude. The prospective proton EDM measurement causes the
distribution of possible electron EDM values to peak more sharply around zero, under either
hypothesis for the proton result, but still allows a similar range of values for the electron EDM.
Discovery of a proton EDM at the level of $\sim 10^{-27}$ e.cm would not preclude the
possibility that the neutron EDM could be an order of magnitude larger, and would not change
radically the prospects for measuring a non-zero value of the electron EDM.

\subsection{CP Asymmetry in $b\to s\gamma$ Decay}

\begin{figure}[t!]
\begin{center}
\includegraphics[width=0.4\paperwidth]{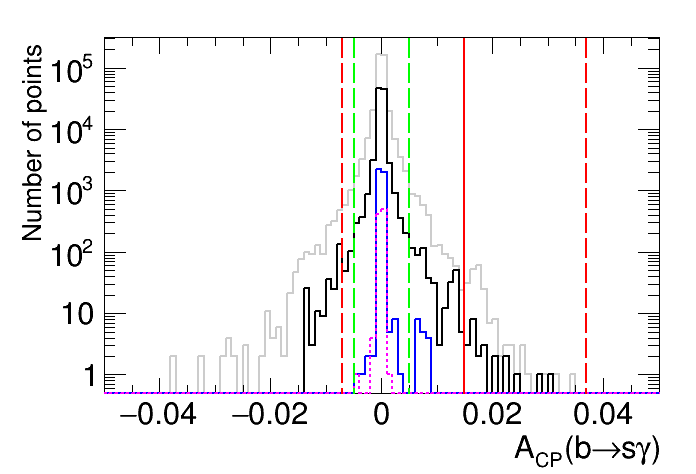}
\caption{\it Distribution of the CP asymmetry in $b\to s\gamma$, $A_{\rm CP}(b \to s \gamma)$. The red dashed line shows the current experimental constraint, while the red solid line indicates its central value, and the green dashed line shows the prospective future sensitivity of Belle-II. Grey histogram: all points passing the LHC mass and flavour constraints; black: also including current EDM limits; blue: including a possible future proton EDM constraint, assuming a central value of zero; magenta: including a possible future proton EDM measurement of $1 \times 10^{-27}$ e.cm, both
with an assumed uncertainty of $\pm 0.025\times10^{-27}$ e.cm.}
\label{ACP}
\end{center}
\end{figure}
\begin{figure}[ht!]
\begin{center}
\includegraphics[width=0.4\paperwidth]{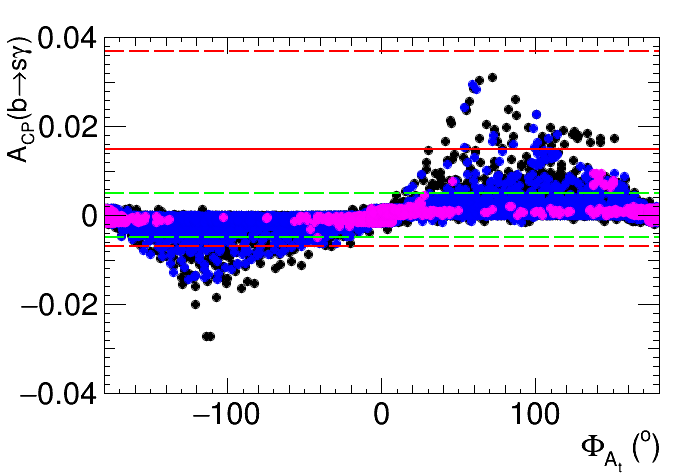}
\caption{\it Distribution of scan points in the $(\phi_{A_t}, A_{\rm CP}(b \to s \gamma))$ plane. The red dashed line shows the current experimental constraint, while the red solid line indicates its central value, and the green dashed line shows the prospective future sensitivity of Belle-II. Black points: all points passing the LHC mass and flavour constraints, as well as current EDM limits; blue: including possible future proton EDM constraint, assuming a central value of zero; magenta:  including a possible future proton EDM measurement of $1 \times 10^{-27}$ e.cm, both
with an assumed uncertainty of $\pm 0.025\times10^{-27}$ e.cm.}
\label{ACP_At}
\end{center}
\end{figure}

Figure~\ref{ACP} illustrates the prospects in the MCPMFV pMSSM for $A_{\rm CP}(b \to s \gamma)$ without (black
histogram) and with (blue and magenta histograms) prospective proton EDM measurements.
The current experimental constraint on $A_{\rm CP}(b \to s \gamma)$ is indicated by the
red solid and dashed lines, and the prospective order-of-magnitude improvement in precision
estimated for Belle-II is indicated by the green dashed lines. We see that the present EDM
constraints are entirely consistent with a positive value of $A_{\rm CP}(b \to s \gamma)$ outside the green dashed lines and hence accessible to Belle-II. On the other hand, the prospective
precision of the proton EDM measurement (whether null or not) would make a Belle-II 
discovery of $A_{\rm CP}(b \to s \gamma)$ seem unlikely. This can be understood from the fact that $A_{\rm CP}(b \to s \gamma)$ is very sensitive to the $A_t$ phase, which appears in chargino-stop loops and will be very much constrained by the future proton EDM measurement. In particular, the current positive $A_{\rm CP}(b \to s \gamma)$ measurement favours positive $A_t$ phases, as demonstrated in Fig.~\ref{ACP_At} where $A_{\rm CP}(b \to s \gamma)$ is plotted as a function of the $A_t$ phase. We observe that a precise measurement of CP asymmetry in $b\to s\gamma$ close to the current experimental central value would induce a large positive $A_t$ phase. This would still be compatible with the current EDM limits as well as a proton EDM limit of $5\times10^{-29}$ e.cm, but would not be compatible with a measurement of proton EDM of, e.g., $(1\pm0.025)\times10^{-27}$ e.cm.

\begin{figure}[t!]
\begin{center}
\begin{tabular}{cc}
\includegraphics[width=0.33\paperwidth]{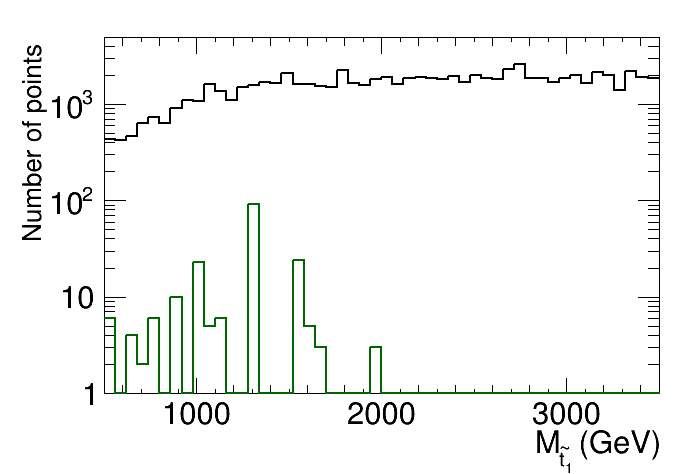} & 
\includegraphics[width=0.33\paperwidth]{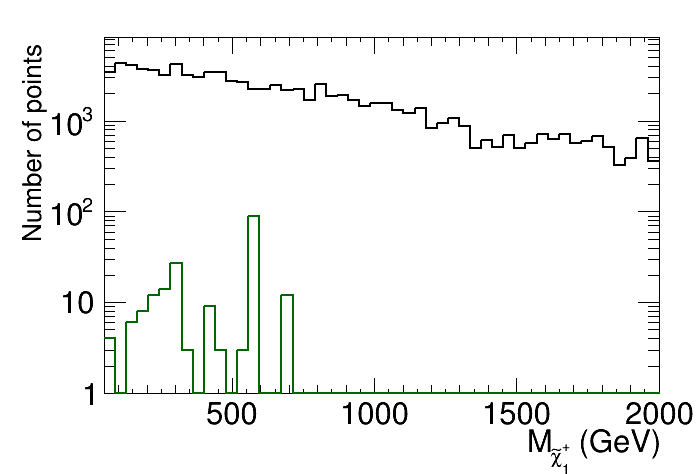}\\
\includegraphics[width=0.33\paperwidth]{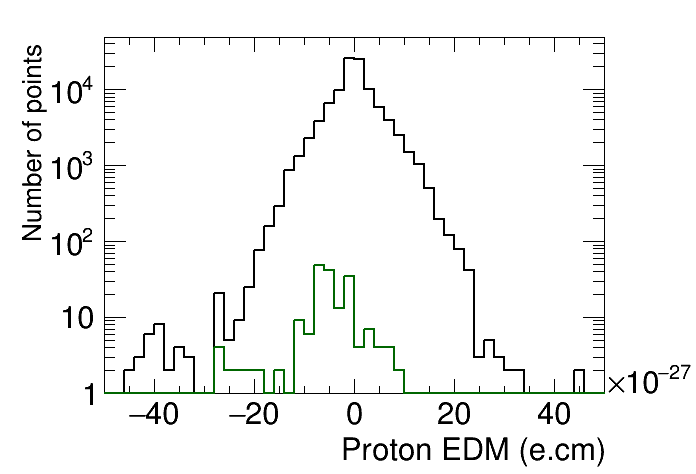} & 
\includegraphics[width=0.33\paperwidth]{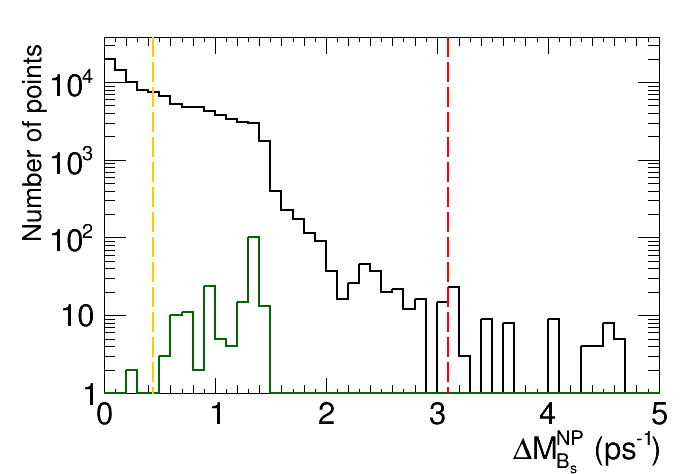}
 \end{tabular}
\caption{\it {Distributions of the lighter stop mass (upper left), lighter chargino mass (upper right), proton EDM (lower left) and $\Delta M_{B_s}^{\rm NP}$ (lower right), showing the effects of a possible Belle-II measurement of $A_{\rm CP}(b \to s \gamma)$ as well as the LHC constraints on sparticle masses and the EDM constraints. Black histograms: all points passing the LHC mass and flavour constraints and the current EDM limits; dark green histograms: including also a possible future measurement {$A_{\rm CP}(b \to s \gamma) = 0.015\pm0.002$}. In the lower right panel we also show the current experimental upper limit on $\Delta M_{B_s}^{\rm NP}$ (vertical dashed red line) and the effect of the prospective reduction in the uncertainty of the relevant hadronic matrix element (vertical dashed yellow line).}}
\label{WhatifACP}
\end{center}
\end{figure}

To complete this Section, we consider the implications of a possible non-null measurement of $A_{\rm CP}(b \to s \gamma)$ by
the Belle-II experiment, with the same central value as the current experimental measurement but a reduced uncertainty of {$0.015\pm0.002$}.
As seen in Fig.~\ref{WhatifACP}, such a measurement would favour (dark green histograms) relatively small values of the lighter stop mass  (upper left panel) and of the lighter chargino (upper right panel), which would be encouraging for future LHC searches. The lower left panel of Fig.~\ref{WhatifACP} shows that such a measurement would leave open the possibility of a proton EDM with a {\it negative} value of magnitude as large as $3 \times 10^{-26}$ e.cm, while positive values would be restricted to $\lesssim 1 \times 10^{-26}$ e.cm, within reach of a prospective proton EDM measurement. Finally, the lower right panel of Fig.~\ref{WhatifACP} shows that such a $A_{\rm CP}(b \to s \gamma)$  measurement would leave open the possibility that $\Delta M_{B_s}^{\rm NP}$ could be as large as $\sim 1.5$~ps$^{-1}$, which could be measured if the uncertainty in the hadronic matrix element could be achieved as expected (vertical dashed yellow line).

\subsection{Contribution to the $B_s - \overline{B}_s$ Mass Difference}

\begin{figure}[t!]
\begin{center}
\includegraphics[width=0.4\paperwidth]{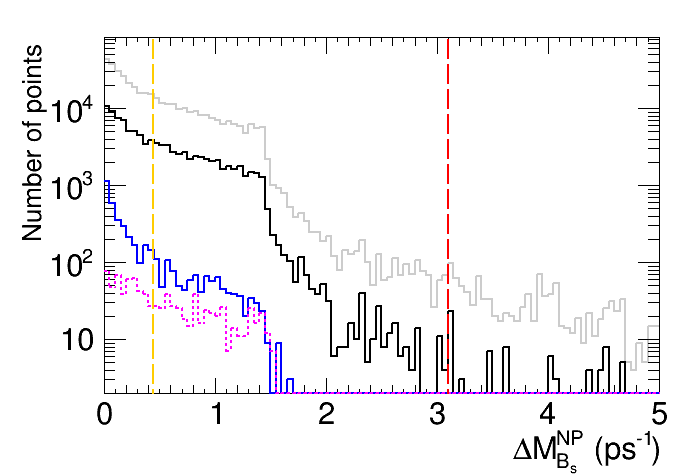}
\caption{\it Distribution of the contribution to the $B_s - \overline{B}_s$ mass difference, $\Delta M_{B_s}^{\rm NP}$. The red dashed line shows the current experimental constraint, and the yellow dashed line shows the prospective future sensitivity assuming that the uncertainty in the non-perturbative calculation can be reduced by a factor of 10. Grey histogram: all points passing the LHC mass and flavour constraints; black: also including current EDM limits; blue: including possible future proton EDM constraint, assuming a central value of zero; magenta: including possible future proton EDM measurement of $(1\pm0.025)\times10^{-27}$ e.cm.}
\label{Bs}
\end{center}
\end{figure}

Figure~\ref{Bs} illustrates the prospects in the MCPMFV pMSSM for a measurable new physics contribution to the
$B_s - \overline{B}_s$ mass difference, $\Delta M_{B_s}^{\rm NP}$. 
The current experimental upper limit on $\Delta M_{B_s}^{\rm NP}$ is indicated by the
red dashed line, and the yellow dashed line indicates the sensitivity that could be gained by
an order-of-magnitude improvement in the precision of the corresponding hadronic
matrix element. We see that the present EDM constraints (black
histogram) allow a value of $\Delta M_{B_s}^{\rm NP} \lesssim 3$~ps$^{-1}$, whereas the
accuracy of the prospective proton EDM measurement would only allow $\Delta M_{B_s}^{\rm NP} \lesssim 2$~ps$^{-1}$, whether the measurement is null, or not. The continuing
prospects in the MCPMFV pMSSM bolster the interest in refining the non-perturbative calculation as far as possible.

\subsection{CP-Violating Higgs Couplings to Third-Generation Fermions}

\begin{figure}[t!]
\begin{center}
\begin{tabular}{cc}
\includegraphics[width=0.33\paperwidth]{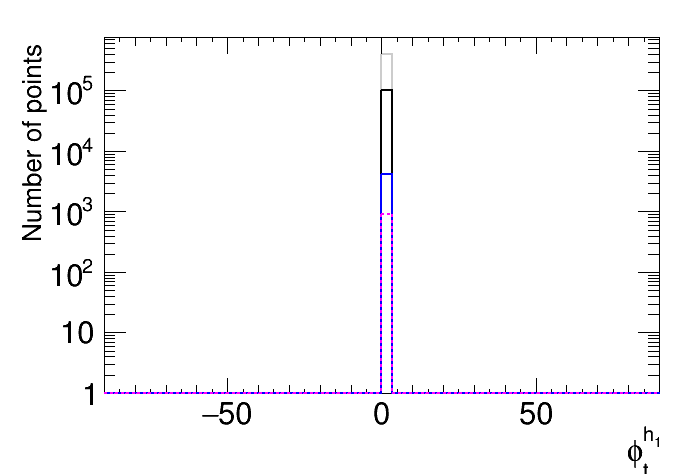} & 
\includegraphics[width=0.33\paperwidth]{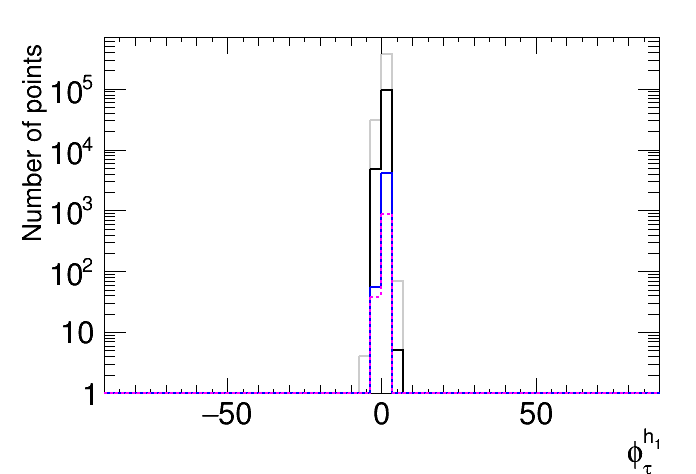} \\
\includegraphics[width=0.33\paperwidth]{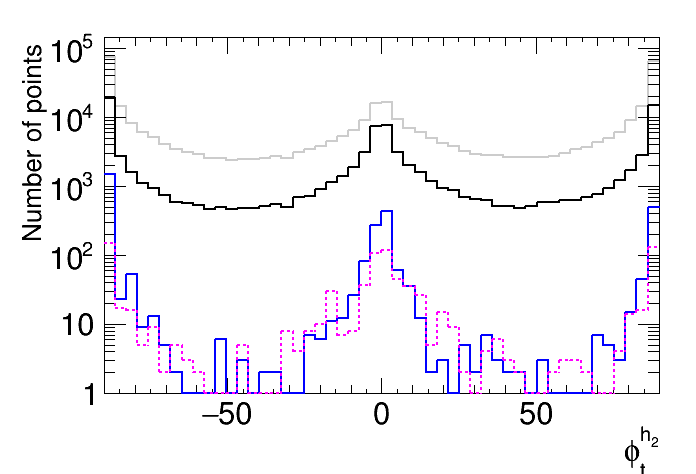} & 
\includegraphics[width=0.33\paperwidth]{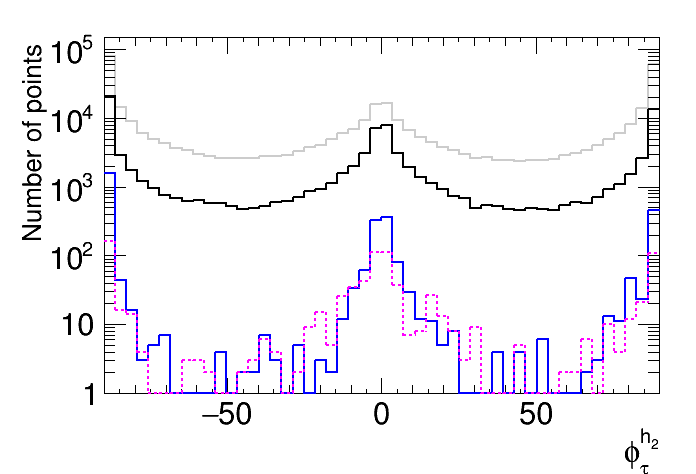}\\
\includegraphics[width=0.33\paperwidth]{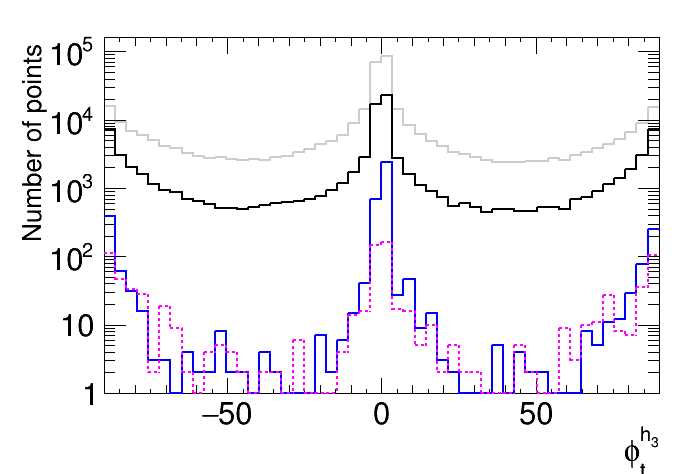} & 
\includegraphics[width=0.33\paperwidth]{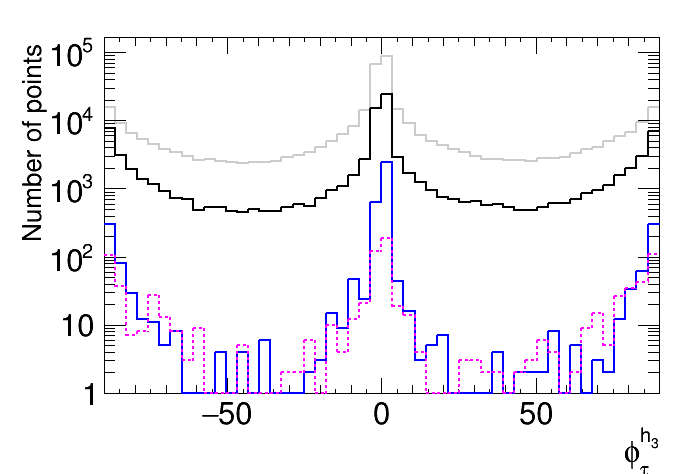}
 \end{tabular}
\caption{\it Distributions of phases of third-generation fermion couplings to neutral Higgs bosons, showing the effects of taking into account the LHC constraints on sparticle masses and imposing the EDM constraints. Left column, from top to bottom: top couplings to $h_{1,2,3}$; right: tau couplings  to $h_{1,2,3}$. Grey histograms: all points passing the LHC mass and flavour constraints; black: also including current EDM limits; blue: including possible future proton EDM constraint, assuming a central value of zero; magenta: including possible future proton EDM measurement of $(1\pm0.025)\times10^{-27}$ e.cm.}
\label{CPh}
\end{center}
\end{figure}

Figure~\ref{CPh} reviews the prospects for measuring CP violation in Higgs-fermion
couplings in the MCPMFV pMSSM. Histograms of the top coupling to the three
neutral MSSM Higgs bosons are shown in the left column, and those of the $\tau$-Higgs
couplings are shown in the right column. It has been suggested that CP violation in the
coupling of the $h(125)$ to the top might be detectable in $t {\bar t} h$ associated
production~\cite{EHST}, but this is already tightly constrained by the upper limit on the electron EDM,
in particular, and the prospective proton EDM hardly changes the situation. Likewise,
CP violation in the coupling of the $h(125)$ to the $\tau$, which could in principle be
measurable via $\tau$ polarization measurements in $h \to \tau {\bar \tau}$ decay~\cite{Berge}, is
also already very tightly constrained. In principle, the present data would allow maximal
CP violation in the $h_{2,3}$ couplings to $t {\bar t} h$ and $\tau {\bar \tau}$, as seen
in the middle and bottom panels of Fig.~\ref{CPh}. These prospects would be
greatly diminished by a precise measurement of the proton EDM, although any of
these couplings might still have a CP-violating phase of ${\cal O}(10^\circ)$.

%

\section{Conclusions}
\label{sec:conclusions}

Everything we know about CP violation is compatible with the Kobayashi-Maskawa mechanism embedded in the
Standard Model. However, we also know that this mechanism is insufficient to explain the cosmological baryon asymmetry,
for which some additional source of CP violation would be required. This might be at the electroweak scale, though this
additional CP violation could well appear at some much higher scale. Upper limits on EDMs constrain significantly
low-scale models of CP violation, but do not exclude them. Supersymmetry is one example of an extension of the
Standard Model at the electroweak scale that contains several CP-violating phases. As emphasized previously and
in this paper, EDM measurements constrain certain combinations of these phases, but leave open the possibility that
other combinations of CP-violating phases might be relatively large. In this case there might be significant deviations
from the predictions of the Kobayashi-Maskawa model for some CP-violating observables.

In this paper we have analyzed the prospects for such deviations in $B$-meson observables
in the framework of the MCPMFV pMSSM~\cite{Ellis:2007kb}, taking into account the
present constraints from unsuccessful searches for sparticles at LEP and during LHC Run~2 as well as the present EDM
constraints and a potential future proton EDM measurement. We have used the geometric approach suggested 
in~\cite{Geometry,Ellis:2010xm}, with the refinements introduced in~\cite{Arbey:2014msa} to sample the MCPMFV 
pMSSM parameter space. We have found that the present EDM constraints are consistent
with values of the CP asymmetry in $b\to s\gamma$, $A_{\rm CP}(b \to s \gamma)$, that are significantly larger than
the the prospective sensitivity of the Belle-II experiment. We have also found that there may be a significant contribution to the
$B_s - \overline{B}_s$ mass difference, $\Delta M_{B_s}^{\rm NP}$, which may be measurable if the current uncertainty
in the calculation of the relevant hadronic matrix element can be reduced significantly. CP-violating phases in the
couplings of the two heavier neutral MSSM Higgs bosons to $\tau$ leptons and top quarks could also be large and
potentially measurable, but not those of the lightest neutral MSSM Higgs boson.

A proposed experiment using a storage ring may be able to measure the proton EDM with an accuracy of 
$0.025\times10^{-27}$ e.cm. Whether this measurement is null or not, it would
leave open the possibility that supersymmetric
CP violation might be measurable in $A_{\rm CP}(b \to s \gamma)$ and/or $\Delta M_{B_s}^{\rm NP}$, 
the latter assuming a plausible reduction in uncertainty in
the calculation of the relevant hadronic matrix element. The cases of the CP-violating phases in the
couplings of the two heavier neutral MSSM Higgs bosons to $\tau$ leptons and top quarks are intermediate: although the
scopes for large phases would be substantially reduced by a precise measurement of the proton EDM, they might
nevertheless be observable. A measurement of the neutron EDM with similar precision to that of the proton EDM would break the 
degeneracies between parameter ranges leading to a small proton EDM.

We have also considered the potential implications of a measurement of $A_{\rm CP}(b \to s \gamma)$ with the present central value
and an uncertainty of {$0.015\pm0.002$}. Intriguingly, we found that this would favour relatively light masses for the lighter stop 
($\lesssim 2$~TeV) and the lighter chargino ($\lesssim 700$~GeV), potentially within the reach of future LHC runs. Moreover, in this case the
proton EDM could well lie within reach of the proposed experiment, and a non-zero value of $\Delta M_{B_s}^{\rm NP}$ could be
inferred if the current hadronic matrix-element uncertainty could be reduced significantly.

Our analysis of the MCPMFV pMSSM shows that, despite the negative results of searches for sparticles during 
Run~2 of the LHC, as well as the stringent upper limits on EDMs, searches for CP violation in $B$ physics, 
in particular, still offer opportunities for making measurements deviating from the predictions of the Kobayashi-Maskawa model. 

\section*{Acknowledgements}
The work of JE was supported in part by the United Kingdom STFC Grant ST/P000258/1, and in part
by the Estonian Research Council via a Mobilitas Pluss grant.


\bibliographystyle{JHEP}

\end{document}